\documentclass{article}
\usepackage{spconf, amsmath, graphicx, float, booktabs, multirow, hyperref}

\begin{document}

\title{Graph2Speak: Improving Speaker Identification using Network Knowledge in Criminal Conversational Data}

\name{Maël Fabien$^{1,2}$, Seyyed Saeed Sarfjoo$^1$, Petr Motlicek$^1$, Srikanth Madikeri$^1$}
\address{
  $^1$Idiap Research Institute, Martigny, Switzerland,\\
  $^2$Ecole Polytechnique Fédérale de Lausanne (EPFL), Switzerland
}


\maketitle

\begin{abstract}
  Criminal investigations mostly rely on the collection of conversational data. The identity of speakers must be assessed in order to build or enrich an existing criminal network. Investigators then use social network analysis tools to identify the most central characters and the different communities within the network. We introduce two candidate datasets for criminal conversational data, Crime Scene Investigation (CSI), a television, and the ROXANNE simulated data. We also introduce the metric of conversation accuracy in the context of criminal investigations. We improve a speaker identification baseline by re-ranking candidate speakers based on the frequency of previous interactions between speakers and the topology of the criminal network, and call this method Graph2Speak.  We show that our approach outperforms the baseline speaker accuracy by 1.3\% absolute (1.5\% relative), and the conversation accuracy by 3.7\% absolute (4.7\% relative) on CSI data, and by 1.1\% absolute (1.2\% relative), and the conversation accuracy by 2\% absolute (2.5\% relative) on the ROXANNE simulated data.
\end{abstract}

\begin{keywords}
criminal conversational data, criminal networks, network analysis, speaker identification
\end{keywords}

\section{Introduction}

Conversational data defines data created by interactions between a set of characters, through text messages, telephone, or video calls for example. In criminal investigations, Law Enforcement Agencies (LEAs) collect criminal conversational data and build criminal networks to assess the links between suspects. Hereby, we introduce ROXANNE~\footnote{\url{https://roxanne-euproject.org/}}, a European Union’s Horizon 2020 research project leveraging real-time network, text, and speaker analytics for combating organized crime.

Speaker identification is a well known tool for criminal investigations, whether used as a tool by investigators \cite{khelif_siip_nodate, morrison_interpol_2016}, or in court \cite{solan_hearing_nodate}. However, real-condition criminal data are hard to collect due to the variety of modalities and channels required. By nature, criminal data are conversational. They require timestamps of all the interactions, names, and roles of each character involved, which are then used by investigators to build criminal networks.

The topology of criminal networks is specific, by the number of characters involved, the number of sub-groups, their sizes, the role of central characters, etc. Moreover, this type of data is private, and only rarely released publicly, and if so, only partial and anonymized data, such as a graph structure, is published \cite{calderoni_robust_2020}.

Criminal networks built from audio data are made of nodes, representing the identity of speakers, and edges, which reflect links between them, all together describing the topology of the network. When audio files are collected, the identity of the characters involved is assessed by a speaker identification system, given the enrolled models from the speakers. Based on the detected identities, a link, i.e. and edge, between two characters in the network is added. Edges can then be weighted to reflect the number of previous interactions between two given characters. Edges with larger weights reflect a high frequency of interactions in the past, which can be crucial information in investigations. 

Criminal networks are typically used for link prediction, i.e. predicting a missing link within a network \cite{calderoni_robust_2020}, as well as node classification, i.e. to identify the class of a node based on some annotated nodes. Social Network Analysis (SNA) is also used to identify communities, central characters \cite{tahalea_central_2019}, or identify the characters to remove from a criminal network \cite{cavallaro_disrupting_2020}. In this paper, we test whether the information contained in the topology of criminal networks can be used to improve speaker identification systems on conversational data, hence favoring strong existing relationships between speakers. 

We propose an extension to existing works by Gao et al.~\cite{lev} by computing, for a conversation, the speaker identification scores of each speaker, and re-ranking the potential speakers based on how frequently they talked to each other over the past and the topology of the criminal network. The main contribution of our approach is that it does not rely on an external source of data, since previous discussions between speakers allow us to build the criminal network, which itself will influence the re-ranking of the next conversations. Our approach also handles more than two speakers per conversation. We introduce a dataset that we used to simulate criminal data, the CSI dataset, and also introduce the metric of the conversation accuracy, which is relevant for LEAs investigations. We reached a relative improvement on the speaker accuracy of 1.5\% on average, and of 4.7\% on average for the conversation accuracy. Our approach shows encouraging results even on heterogenous datasets in which some speakers spoke for more than 6 minutes, and others for 20 seconds only.

Section \ref{sec:data} presents the Crime Scene Investigation (CSI) data as a potential candidate for criminal conversational data. The evaluation metrics used are described in Section \ref{sec:metrics}. The results of a baseline speaker identification are presented in Section \ref{sec:baseline}, while section \ref{sec:method} describes the existing re-ranking method by Gao et al.~\cite{lev}, as well as our method and experimental results. Section \ref{sec:discuss} discusses obtained results, the dataset, and the metrics used, as well as the future direction of network-based improvement of speaker identification. Finally, Section \ref{sec:concl} depicts our conclusions.

\section{Related work}

The idea of improving the speaker identification process using an external source of data is not new. It has been explored by Khelif et al. in \cite{khelif_siip_nodate}, by operating an inter-task fusion, using the accent, the language or the gender as an external source of information. In \cite{madikeri_bayesian_2019}, Madikeri et al. obtained a relative improvement of the Equal-Error-Rate (EER) of 8\% using a probabilistic fusion on NIST SRE 2008.

Gao and al.~\cite{lev} have shown in previous works on Enron email and phone call databases that we can re-rank speaker pairs using network information. The knowledge present in the email database was used to build a network, and assess how often speakers talked to each other. This information was then used to re-compute the score of a pair of speakers, improving the score of the pair if speakers talked to each other frequently in the past. This work has shown an improvement in classification error and on the harmonic mean of the rank of the known speaker. This approach requires an external source of data, such as emails in the case of Enron, and focuses on the sub-case of conversations between 2 characters.

Due to the lack of coherent data and the amount of data preparation needed, only a few works have explored the impact of network analysis in speaker identification for criminal conversational data.

\section{Criminal Conversational Data}
\label{sec:data}

To the best of our knowledge, apart from the Enron e-mail database~\cite{hutchison_enron_2004} augmented with the Enron phone call database \footnote{\url{https://web.archive.org/web/20060208051051/
http://www.enrontapes.com/files.html}}, as described by Gao et al.~\cite{lev}, no real-condition criminal conversational database exists. This dataset is a subset of conversations between managers at Enron, a company which was convinced of corporate fraud in 2001. However, most fraudulent conversations were removed from Enron database, and the topology of the network we can build does not entirely reflect the fraudulent activities of Enron. 

\subsection{CSI dataset}

We propose to use CSI television show as a potential candidate for criminal investigation data. CSI is a popular criminal investigation television series in the United-States. Each episode of the series includes a video of around 40 minutes, an audio file, and a transcript. The audio and video are extracted from the DVD of the show. The transcripts were published by the natural language processing group of the University of Edinburgh for previous work on automatic killer identification in CSI episodes \cite{whodunnit}. The transcripts also describe the role of each speaker (suspect, killer, or other) and can be downloaded publicly on GitHub \footnote{ \url{https:\/\/github.com\/EdinburghNLP\/csi-corpus} }. Each episode involves a team of investigators, journalists, victims, the family of the victims, suspects, and killers.

\begin{figure}[H]
  \centering
  \includegraphics[width=\linewidth]{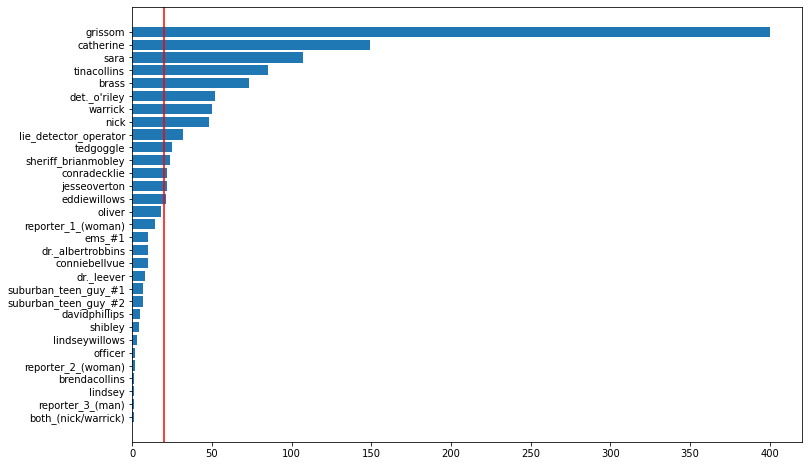}
  \caption{Distribution of aggregated speech time, in seconds, per speaker in season 1 episode 7.}
  \label{fig:dist}
\end{figure}

We collected transcripts of 39 episodes and video/audio of 4 episodes. Each episode involves on average more than 30 speakers. Utterances last on average 3 to 4 seconds. There are around 45 to 50 distinct scenes/conversations per episode. The timestamps of conversations in all episodes presented below were identified by hand. Figure \ref{fig:dist} presents the distribution of the speech duration per speaker in season 1 episode 7. 

Although the episodes of CSI focus on the investigators as well as suspects, whereas a real investigation would only collect data on suspects, we suppose that the structure of the networks that we can extract from this dataset is relevant for criminal investigation, by the number of speakers, the frequency of the interactions between the speakers, the number of sub-groups in each episode, the role of central characters who act as information bottlenecks... The topics of the conversations and the vocabulary used will also allow future works on Automatic Speech Recognition (ASR) systems in the context of criminal conversations. The episodes sequentially display a murder, the body/bodies are then discovered, police start the investigation, gather evidence, interrogate suspects, and identify the killer. We do not have information on the exact time at which each scene took place. Therefore, we consider the data as sequential by default and do not conduct any analysis on the time between conversations.

\begin{figure}[!t]
  \centering
  \includegraphics[width=\linewidth]{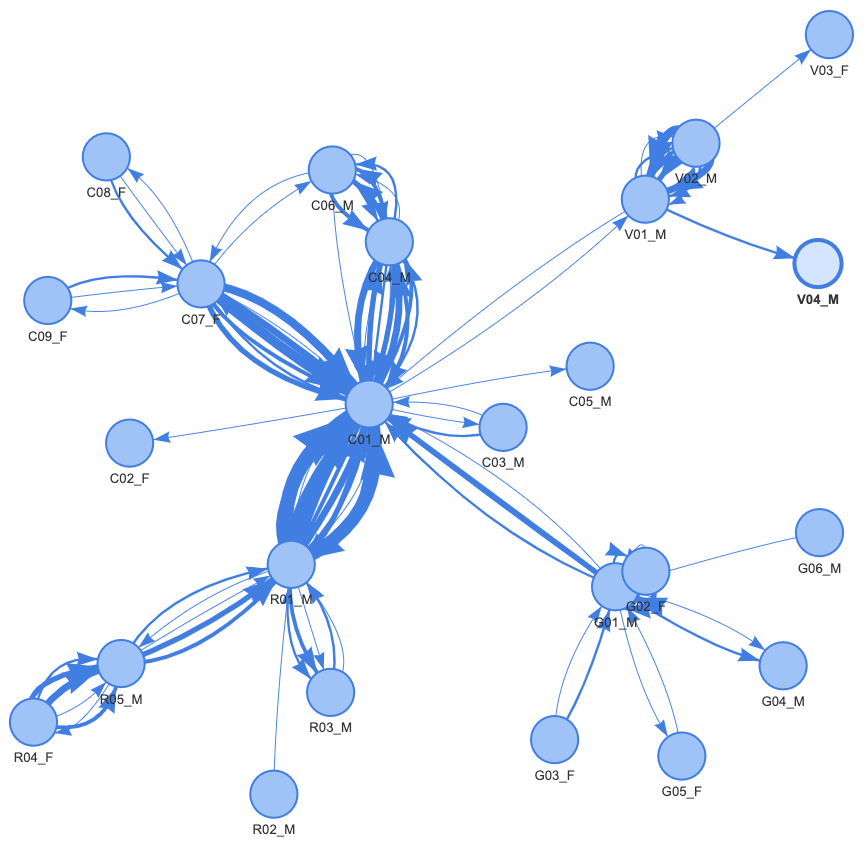}
  \caption{Network representation of the simulated data.}
  \label{fig:net}
\end{figure}

We build the "ground-truth" network using the transcripts provided, as illustrated in Figure \ref{fig:net}. In the interactive tool developed, the thickness of the edges reflects the number of interactions between the speakers, and the tool-tip displayed on the node shows the name of the character. Note that the network we built is only a sub-network from the whole Season 1 Episode 7, focusing on the 14 characters with more than 20 seconds of speech. Speakers without sufficient data were mostly journalists or local police officers, and did not have a central role in the conversations.

Like real investigations, CSI data is subject to changes of the frequency of interactions between characters, if for example a character tends to become more and more important over the course of the episode. However, CSI data might be harder than real investigation data in the sense that over the span of 3 or 4 conversations, a character might completely disappear from the episode, whereas this type of event typically happens in real investigation over the course of more interactions between characters, which leads to smaller changes in the network topology.

\subsection{ROXANNE simulated dataset}

In order to test our approach on more realistic data, the ROXANNE consortium has also collected simulated data. A screenplay was carefully prepared by one of the LEA partners, and the phone calls were made by members of the consortium, and recorded using Twilio. The simulated dataset is made of 24 speakers, who made 100 phone calls, resulting in 155 minutes of speech, on topics related to drugs, money and meetings, in Czech, Russian, Vietnamese, German and English. This dataset, although acted, provides close-to-real-life conditions. The topology of the network is meant to reflect the structure of a real-life criminal network. The multilingual framework, the background noises and the short utterances present in the dataset are also challenging elements which match real cases. 

\section{Evaluation metrics}
\label{sec:metrics}

Speaker accuracy is a natural candidate metric for a speaker identification task in criminal investigations. Another metric which is relevant for LEAs is the percentage of conversations for which we could identify all speakers, which we define as the conversation accuracy. In a criminal case with $C$ conversations, each of the conversations involves the list of speakers $s_c$. Using our speaker identification system, we predict the list of speakers as being $s_{pc}$. The conversation accuracy then becomes:

\begin{equation}
acc_C =\frac{1}{C} \sum_{c=1}^{C} \delta{(s_{pc}, s_c)},
\end{equation}
where $ \delta{(s_{pc}, s_c)} $ is an indicator function equal to 1 if $s_{pc}$ is equal to $s_c$.

\begin{figure}[!t]
  \centering
  \includegraphics[width=\linewidth]{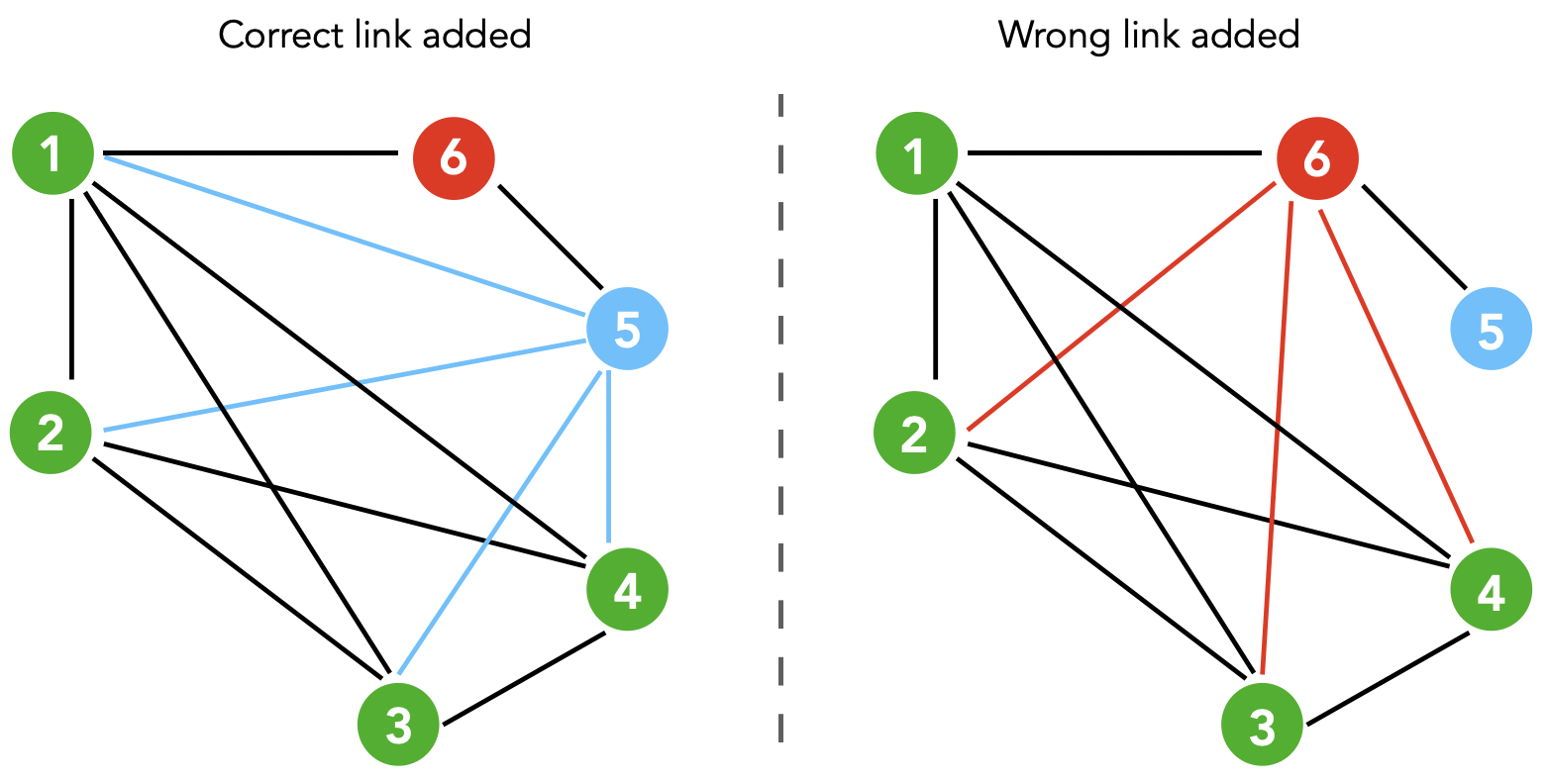}
  \caption{Effect of adding a wrong edge to an existing network.}
  \label{fig:wrong}
\end{figure}

The motivation behind the conversation accuracy is that adding a wrong edge to the network of known connections could lead investigators on the wrong track. Let us illustrate the process of misclassifying one of the speakers in a conversation in Figure \ref{fig:wrong}, based on an existing network. The existing links between speakers are presented in black. In a new conversation, we suppose that speakers 1, 2, 3, 4 and 5 were talking. On the left part of Figure \ref{fig:wrong}, we added the correct edges in blue. On the contrary, if we misclassify speaker 5 as being speaker 6, three wrong edges are added at once, and the topology of the network is completely modified, which motivates the use of this metric in criminal investigations. 

Such a metric has the advantage of being easy to interpret for LEAs, easy to implement, and close to the objective LEAs pursue of building accurate and reliable criminal networks at the conversation level. To compute it, we split the raw audio file of each episode into a sequence of conversations. These sequences of conversations were manually annotated for each of the 4 episodes of CSI, but could also be inferred automatically using speaker diarization techniques~\cite{diar}. The identity of each speaker in a conversation is then assessed, and after processing all the conversations, the conversation accuracy is computed.

\section{Speaker identification baseline}
\label{sec:baseline}

Due to the relatively low volume of data available, we used a pre-trained speaker identification system prepared for the NIST Speaker Recognition Evaluation (SRE19) dataset~\cite{sadjadi_2019_nodate}. The pre-trained system is described in Idiap's submission to the NIST SRE 2019 Speaker Recognition Evaluation~\cite{Sarfjoo_Idiap-RR-15-2019}. The submission relies on Time Delay Neural Network (TDNN)~\cite{peddinti_time_nodate} X-vector systems~\cite{xvec, snyder_speaker_2019} with a Probabilistic Linear Discriminant Analysis (PLDA)~\cite{leonardis_probabilistic_2006} back-end. 

We first downsampled speech data to 8\,kHz (with an application of band-pass filtering between 20 and 3700\,Hz). Then, 23-dimensional mel frequency cepstral coefficients (MFCCs) were extracted on 25\,ms speech windows, with a frame-shift of 10\,ms. To remove non-speech frames, energy-based Voice Activity Detection (VAD) was applied. 

We trained the X-vector system on Voxceleb dataset~\cite{nagrani_voxceleb_2018} and on the augmented versions of Switchboard dataset~\cite{godfrey_switchboard_1992} and SRE 2004 to 2010 with additive noise (MUSAN dataset~\cite{snyder_musan_2015}) and reverberation (RIR dataset~\cite{szoke_building_2019}). The PLDA classifiers were trained on augmented versions of SRE. 

For CSI dataset, we only selected speakers for which we were able to collect at least 20 seconds of audio samples. If we were able to collect at least 40 seconds of speech for a character, we kept 20 seconds as enrolment and the rest in test. Depending on the episode, we have 13 to 15 speakers among a total of 28 to 33 speakers. The X-vector/PLDA baseline heads an average speaker accuracy for the close-set speaker identification task of 89.9\% on the 4 episodes and an average conversation accuracy of 78.1\%.

For the simulated data, we took the first utterance of each speaker as en enrollment file, and the rest in test. We therefore enrolled 24 speakers. The X-vector/PLDA baseline reached a speaker accuracy of 88.1\%, and a conversation accuracy of 79.6\%.

\section{Graph2Speak Re-ranking algorithm}
\label{sec:method}

\subsection{Method description}

We introduce $s_{mc}$ as being a joint score of all speakers in a conversation $c$, considering the combination of speakers $m$. Our aim is to score all combinations of speakers ($M_c$ in total) in conversation $c$, and choose the combination which maximizes the score. In the given conversation, there are $N_{mc}$ different speakers. For each speaker $k$, we obtain the acoustic score $s_k$ from the X-vector baseline. We define the relative degree centrality $C_k$ as the number of interactions of speaker $k$ divided by the total number of interactions in the network at that moment of the conversation $c$, denoted $E_c$. The score of the combination $m$ of speaker for conversation $c$ can be written as:

\begin{equation}
s_{mc}=\frac{\sum_{k=1}^{N_{mc}} s_k (1 + C_k)}{N_{mc}}  \prod_{i < j}^{N_{mc}}{(1+\lambda \frac{e_{i,j}}{E_c})}
\end{equation}

The second part of the score computes all the permutations of speakers, two-by-two, denoted $i$ and $j$, within the list of candidates $m$, $e_{i,j}$ is the number of times speakers $i$ and $j$ talked to each other over the past. The factor $\lambda$ denotes a weighting factor, which we set to 1 by default, but has been adjusted to 0.2 in some of our experiments. 

The logic behind this scoring approach is to weight the acoustic scores of each speaker by their relative degree centrality to favor speakers who have talked much over the past. Then, we multiply the resulting score by the frequency of the conversations between all the permutations of the speakers. For example, if characters A, B, and C talked frequently over the past, then the two-by-two permutations between A and B, B and C, A and C will lead to a large increase of the score of this combination of candidates. For a given conversation $c$, we will select the optimal combination of speakers $m$ such that:
\begin{equation}
s_{mc}^* = {arg\,max}_{m \in M_c} s_{mc}
\end{equation}

\begin{figure*}[!t]
  \centering
  \includegraphics[width=\linewidth]{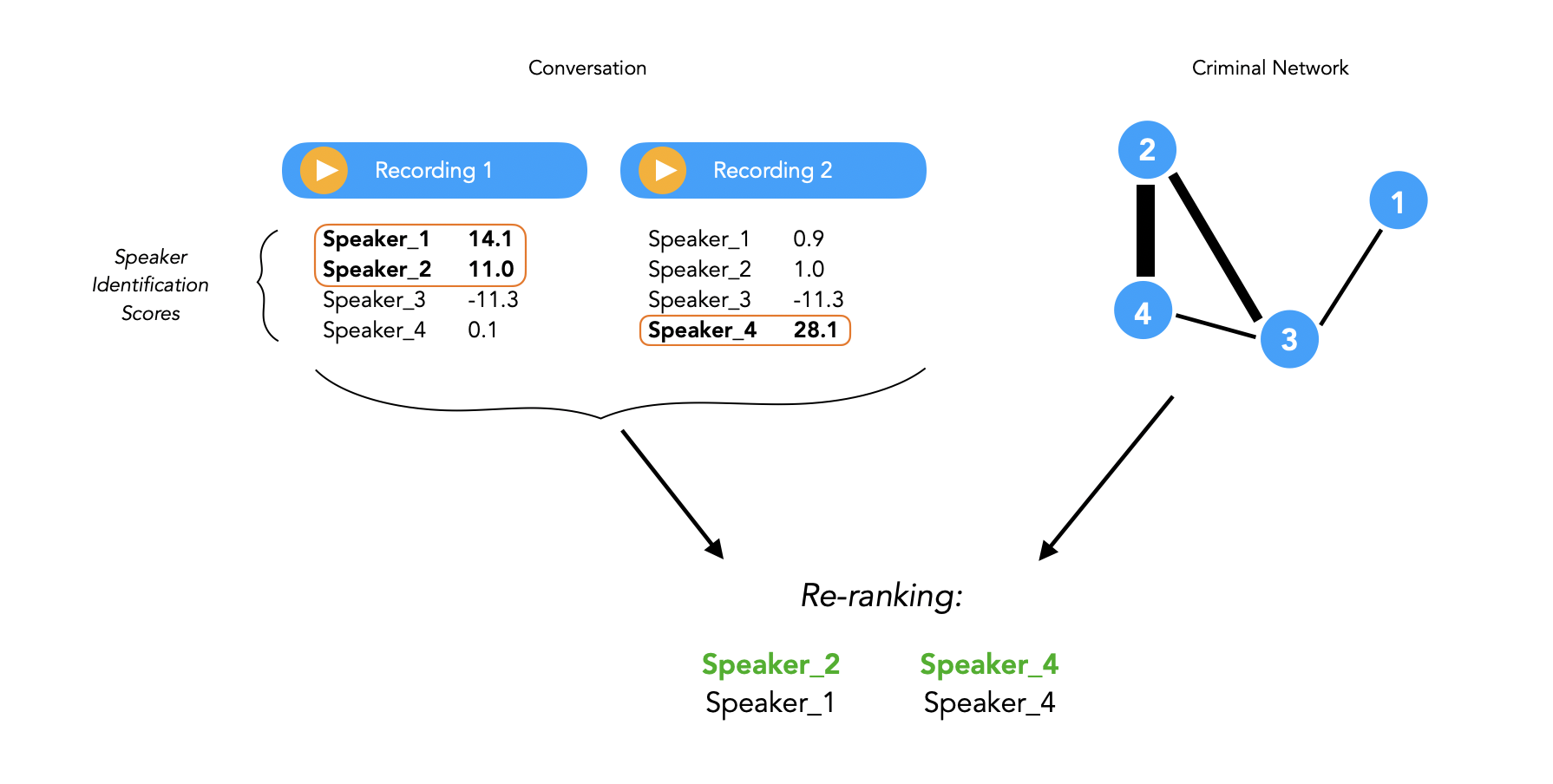}
  \caption{Graph2Speak re-ranking process}
  \label{fig:rank}
\end{figure*}

The process of our method is presented in Figure \ref{fig:rank}. We can notice that for the first recording in the conversation, Speaker 1 and Speaker 2 have acoustic scores from the speaker identification system which are relatively high, and close. For the second recording, Speaker 4 is by far the candidate with the highest score. However, from the topology of the network, we see that Speakers 2 and 4 have been talking a lot over the past and Speaker 1 and 4 never spoke together. Through the re-ranking process, we multiply the score of Speaker 2 by its relative degree of centrality and the score of Speaker 4 by its degree of centrality. We then average the two scores and multiply the result by the relative number of interactions between Speaker 2 and 4. The score we obtain for the pair of Speaker 2 and 4 is higher than the pair of Speaker 1 and 4. Therefore, the re-ranking favors speakers with a high frequency of interactions in the past. 

The novelty of our approach compared to~\cite{lev} is to focus on a single data source, and not external ones. The re-rankings that we operate have impacts on the ways we built the network. We then estimate the number of interactions and the centrality of characters on this network, which will itself influence the next re-ranking.

Since some of the computations imply evaluating all combinations between all speakers involved in a conversation, it can create a large number of combinations to compute. In order to limit the number of combinations tested, we apply a threshold on the scores under which we decide not to consider a speaker as a potential candidate for a given recording as part of a conversation.

\subsection{Experimental results}

For CSI dataset, we compute the scores on 6 episodes, resulting in 5 hours of conversations. For each episode, instead of re-using the same network, we created a new one. The X-vector baseline correctly identified all speakers in 78.1\% of the conversations on average. Using the Graph2Speak approach, conversation accuracy reaches 81.8\%. Speaker accuracy has also been improved from 89.9\% to 91.25\%. 

For the simulated data, the Graph2Speak approach improved the speaker accuracy by 1.1\% (absolute), at 89.2\%, and the conversation accuracy by 2\% (absolute), at 81.6\%. 

A summary of the results is presented in Table \ref{tab:res}. 

\begin{table*}[!t]
  \caption{Graph2Speak performance summary on CSI episodes, in percentage.}
  \label{tab:word_styles}
  \centering
  \begin{tabular}{l || c | c || c | c}
    \toprule
    & \multicolumn{2}{c|}{\textbf{Speaker Accuracy (\%)}} & \multicolumn{2}{c|}{\textbf{Conversation Accuracy (\%)}} \\
    Source & Baseline & Graph2Speak & Baseline  & Graph2Speak \\
    \midrule
    CSI S01E07 & 91.6 & {\fontseries{b}\selectfont 92.7} & 84.4 & {\fontseries{b}\selectfont 88.8}            \\
    CSI S01E08  & 91.9 & {\fontseries{b}\selectfont 95.3} & 80.6 & {\fontseries{b}\selectfont 88.8}                  \\
    CSI S02E01 & {\fontseries{b}\selectfont 88.0} & {\fontseries{b}\selectfont 88.0} & 71.4 & {\fontseries{b}\selectfont 73.5}                 \\
    CSI S02E04  & 88.1 & {\fontseries{b}\selectfont 89.0} & {\fontseries{b}\selectfont 76.1} & {\fontseries{b}\selectfont 76.1}               \\
    CSI S02E06  & 86.2 & {\fontseries{b}\selectfont 88.6 } & 70.9 & {\fontseries{b}\selectfont 74.5}               \\
    CSI S02E09 & {\fontseries{b}\selectfont 92.3} & 91.3 & {\fontseries{b}\selectfont 81.4} & 79.1    \\
    \midrule
    CSI Average  & 89.5 & {\fontseries{b}\selectfont 90.6} & 77.7 & {\fontseries{b}\selectfont 80.2}               \\
    \midrule
    \toprule   
    Simulated data & 88.1 & {\fontseries{b}\selectfont 89.2} & 79.6 & {\fontseries{b}\selectfont 81.6}             \\
    \midrule
    \bottomrule
  \end{tabular}
  \label{tab:res}
  
\end{table*}

On CSI, we reached a relative improvement of 4.7\% in terms of conversation accuracy and 1.5\% in speaker accuracy. For conversation and speaker accuracy, we obtained absolute improvements of 3.7\% and 1.3\%, respectively. On the ROXANNE simulated data, the relative improvements reach  2.5\% and 1.2\% respectively.

In CSI, the way teams work on investigations is usually structured. The members of the investigation police are split into groups, and each group works on specific tasks. Most conversations hence take place within rather small groups. In this case, the network topology reflects the structure of groups working on a case, and our approach improves the speaker accuracy and the conversation accuracy by a significant factor. This is typically the situation with slow changes in network topology that we expect to encounter in real investigations. In other cases, the whole team works on the investigation without any distinct group being made, e.g. in S02E01, and the frequency of the conversations between speakers is constantly modified. In that case, speaker accuracy is not improved, but we do still observe a slight improvement in conversation accuracy. Note that in this episode, the baseline X-vector approach performs quite poorly since the episode takes place in a casino, with a lot of background noises.

S02E09 also offers an interesting scenario. Two distinct cases are at first investigated by two distinct groups. Then, as the investigation moves on, both murders appear to be related, and the teams working on the cases are mixed again. This illustrates limitations of the approach in networks which quickly change structure. 

\subsection{Performance illustration}

\begin{figure}[H]
  \centering
  \includegraphics[width=\linewidth]{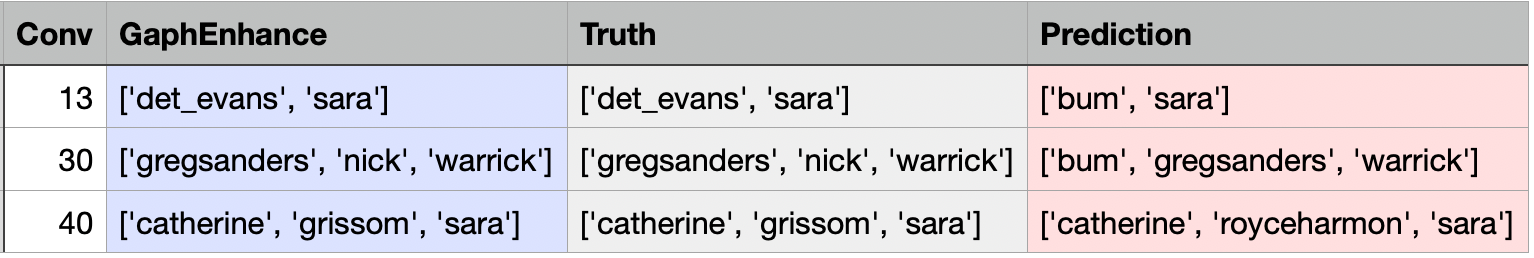}
  \caption{Speaker identification results in conversations where Graph2Speak outputs something different from the baseline}
  \label{fig:diff}
\end{figure}

For the episode 8 of season 1, which is largely improved by the Graph2Speak approach, we selected only the predicted conversations for which different speaker were predicted between the baseline Speaker Identification and our model. Figure \ref{fig:diff} displays such cases. The ground truth is the column in grey, the Graph2Speak output is in blue, and the baseline speaker identification is in red.

In this episode, the 2 differences lead to a correct re-ranking of authors, hence avoiding wrong links to be added in the final criminal network. 

\section{Discussions}
\label{sec:discuss}

In this paper, we first present the CSI dataset as a potential candidate for criminal investigation data. Although some characters of the episodes are the members of the investigation police, CSI remains a good potential candidate for criminal conversational data by the number of characters, the frequency of the conversations between key characters, the number of sub-groups, the information bottlenecks... The network is time-varying, meaning that characters are discovered gradually and new interactions take place sequentially. The topology of the network reflects the creation of various groups (investigation, suspects, ...). The identification of the central speakers is also coherent since the main investigators also appear as the most central characters in our analysis. CSI is however not a perfect dataset yet. Conversations would need to be recorded over the telephone, and being less acted. Additional data could be provided, such as emails and SMS between characters. Further work should be conducted in order to build a new candidate dataset for criminal investigations. 

We also introduce the ROXANNE simulated dataset, a collection of phone calls between members of the ROXANNE consortium, with a screenplay prepared by LEAs. This dataset respects the main characteristics of a criminal investigation (recorded phone calls, network structure, discussed topic, length of conversations...), and will be further used in ROXANNE project as a baseline to evaluate various models.

Graph2Speak, the proposed re-ranking approach, explores the different permutations of speakers for each conversation. There are three novel elements in our approach compared to~ \cite{lev}. We do not rely on an external source of data to estimate the number of links between speakers, but exclusively on previous interactions in the same data source. Therefore, re-ranking decisions made previously impacts the criminal network we build, which impacts the number of links between 2 candidate speakers in a conversation later on. We then offer an extension by applying our approach to more than 2 speakers in a conversation, thus creating large combinatorial factors controlled by thresholds. Finally, we apply our method in the context of a criminal investigation and show interesting improvements in speaker and conversation accuracy. Our approach shows encouraging results even on heterogenous datasets in which some speakers spoke for more than 6 minutes, and others for 20 seconds only, and keeps improving the performance even in cases where the frequency of conversations changes rapidly.

We have shown a significant accuracy gain in the context of criminal investigations, on CSI data and on the ROXANNE simulated data. We conducted experiments in the context of the ROXANNE project and focused on criminal conversational data. Conversational data is broader than criminal investigations, and we do expect that one can improve speaker identification systems in other contexts, on larger volumes of data and wider networks. 

The re-ranking approach that we propose remains simple, and several types of fusion could be also explored in future works. Network attributes, other than relative degree centrality and number of edges between two characters, could also be leveraged. We could indeed include notions from community detection or hierarchical embedding in the re-ranking algorithm. 

\section{Conclusions}
\label{sec:concl}

We introduced both CSI dataset and the ROXANNE simulated data as potential candidates for criminal investigation data. We also introduced the metric of conversation accuracy, and have shown that our re-ranking method based on previous interactions can improve speaker identification on CSI dataset by a relative 1.5\%, and conversation accuracy by 4.7\%, on on the ROXANNE simulated data, by 1.2 and 2.5\% respectively. 

We discussed the limits of these datasets and of our re-ranking method, and offer some future directions to take by including social network analysis tools in speaker identification.

\section{Acknowledgment}

This work was supported by the European Union’s Horizon 2020 research and innovation program under grant agreement No. 833635 (project ROXANNE: Real-time network, text, and speaker analytics for combating organized crime, 2019-2022).

\bibliographystyle{IEEEbib}

\bibliography{bib}

\end{document}